\documentclass[conference]{IEEEtran}
\IEEEoverridecommandlockouts
\newcommand{\vect}[1]{\boldsymbol{#1}}
\def\endthebibliography{%
  \def\@noitemerr{\@latex@warning{Empty `thebibliography' environment}}%
  \endlist
}
\makeatletter
\def\ps@IEEEtitlepagestyle{%
  \def\@oddfoot{\mycopyrightnotice}%
  \def\@evenfoot{}%
}
\def\mycopyrightnotice{%
  {\footnotesize 978-1-7281-8192-9/21/\$31.00 ©2021 IEEE\hfill}% <--- Change here
  \gdef\mycopyrightnotice{}% just in case
}
\usepackage{hyperref}
\usepackage{amsmath,amssymb,amsfonts}
\usepackage{algorithmic}
\usepackage{graphicx}
\usepackage{breqn}
\usepackage{enumitem}
\usepackage{array}
\newcolumntype{P}[1]{>{\centering\arraybackslash}p{#1}}
\usepackage{textcomp}
\usepackage{xcolor}
\usepackage{lipsum}% http://ctan.org/pkg/lipsum
\usepackage{lettrine}
\def\BibTeX{{\rm B\kern-.05em{\sc i\kern-.025em b}\kern-.08em
    T\kern-.1667em\lower.7ex\hbox{E}\kern-.125emX}}
\begin{document}
\title{Probabilistic Voltage Sensitivity based Preemptive Voltage Monitoring in Unbalanced Distribution Networks\\}
\author{\IEEEauthorblockN{Mohammad Abujubbeh, ~\textit{Student Member, IEEE}, Sai Munikoti, ~\textit{Student Member, IEEE}, \\ Balasubramaniam Natarajan,~\textit{Senior Member, IEEE}
\thanks{M. Abujubbeh, S. Munikoti, and B. Natarajan are with Electrical and Computer Engineering, Kansas State University, Manhattan, KS-66506, USA, (e-mail: abujubbeh@ksu.edu, saimunikoti@ksu.edu, bala@ksu.edu)
}}}

\maketitle

\begin{abstract}
With increasing penetration of renewable energy and active consumers, control and management of power distribution networks has become challenging. Renewable energy sources can cause random voltage fluctuations as their output power depends on weather conditions. Conventional voltage control schemes such as tap changers and capacitor banks lack the foresight required to quickly alleviate voltage violations. Thus, there is an urgent need for effective approaches for predicting and mitigating voltage violations as a result of random fluctuations in power injections. This work proposes a novel voltage monitoring approach based on low-complexity, data-driven probabilistic voltage sensitivity analysis. The usefulness of this work is not only in predicting voltage violations in unbalanced distribution grids, but also in opening up the door for optimal voltage control. Using system data and forecasts, the proposed approach predicts the distribution of system node voltages which is then used to to identify nodes that may violate the nominal operational limits with high probability. The method is tested on the IEEE 37 node distribution system considering integrated distributed solar energy sources. The method is validated against the classic load flow based method and offers over 95\% accuracy in predicting voltage violations. 
\end{abstract}
\section{Introduction}
The integration of smart grid technologies such as electric vehicles, energy storage facilities, and distributed generation, introduces advantages as well as system operational challenges \cite{al2019iot}. Renewable energy sources are characterized by variable power outputs that increase system vulnerability to operational inefficiencies \cite{abujubbeh2019determining}. In particular, distribution grids become highly vulnerable to random voltage fluctuations especially when there is a high penetration of distributed solar PV generation \cite{jhala2019data} \cite{jhala2018probabilistic}. Conventional voltage regulation methods such as capacitor banks \cite{madruga2010allocation} and tap changers \cite{zad2013coordinated} represent reactionary approaches and do not exploit any knowledge of voltage state based on anticipated power fluctuations. One reason for resorting to such reactive approaches is the difficulty in estimating the states of a distribution network due to lack of observability. However, recent efforts on sparsity-based estimation strategies (see \cite{jhala2017probabilistic,Munikoti2020probabilistic,jhala2019dominant}) have opened up new possibilities for more proactive methods for voltage regulation \cite{jhala2019data}. Additionally, the classic voltage regulation methods are not designed for bi-directional current flow and typically provide reactive support after an event is detected \cite{seguin2016high}. Many recent research efforts have explored the possibility of using reactive power capabilities of PV generators through smart inverters in either a centralized \cite{deshmukh2012voltage}, \cite{barr2014integration} or decentralized \cite{demirok2011local} scheme. The efficacy of these methods is dependent on the ability to accurately predict voltage violations in the system so that operational setpoints of the PV inverters can be appropriately set in advance. Load flow based look-ahead prediction approaches are cumbersome, computationally complex and not scalable. Therefore, the development of a computationally efficient, yet accurate voltage-violation prediction approach that predicts future violations as well as their concomitant uncertainty bounds is critically important for control and management of distribution grids. This paper aims at developing and testing a computationally efficient voltage violation prediction scheme while considering different penetration levels of PV generation. Based on our prior work on probabilistic voltage sensitivity analysis (PVSA) \cite{jhala2017probabilistic}, \cite{8973956}, the present research focuses on identifying nodes with high probability of violating voltage limits at different time instances. Leveraging existing knowledge of voltage states along with uncertain forecasts of power generation/consumption, probabilistic voltage sensitivity analysis is used to reveal impending voltage issues at any node in the network.  The major scientific contributions of this work include:
\begin{itemize}
\item A computationally efficient, analytical approach to compute the probability of voltage change at any node in an unbalanced distribution system as a result of change in real and reactive power injections at multiple active consumer locations is proposed.
\item A probabilistic voltage sensitivity analysis based approach that predicts the probability of future voltage violations due to change in complex power injection is developed. The approach is used to predict the number of violations in the system at any time instant (based on forecasted PV generation). 
\item The complexity of the proposed analytical approach is significantly lower than traditional load flow-based methods. 
\end{itemize}
\section{Background: Voltage Sensitivity Analysis}
\label{section:background}
For a given three-phase distribution network, analytical voltage sensitivity analysis estimates the complex voltage change at a particular node (observation node $O$) as a result of complex power change at another node (actor node $A$) in the system \cite{8973956}. The usefulness of this approach is seen in the reduced computational complexity in comparison with Newton-Raphson based power flow methods. The change of power consumption at an actor node $A$ from $S_A$ to $S_A + \Delta S_A$ results in voltage change at observation node $O$ from $V_O$ to $V_O + \Delta V_O$. The voltage sensitivity for a given observation node $O$ can be calculated using theorem 1 \cite{8973956}.\\

\textbf{Theorem 1.} For a given three phase distribution network, the change in voltage at an observation node ($\vect{\Delta V_O}$) due to change in power consumption at an actor node ($\Delta S_A$) is approximated by:
\begin{equation}  
 \vect{\Delta V_O} \approx   - \begin{bmatrix} \frac{ \Delta S_{A}^a Z^{aa}_{OA}}{V^{a*}_{A}}+\frac{ \Delta S_{A}^b Z^{ab}_{OA}}{V^{b*}_{A}} + \frac{ \Delta S_{A}^c Z^{ac}_{OA}}{V^{c*}_{A}} \\
 
                               \frac{ \Delta S_{A}^b Z^{ba}_{OA}}{V^{a*}_{A}}+\frac{ \Delta S_{A}^b Z^{bb}_{OA}}{V^{b*}_{A}} + \frac{ \Delta S_{A}^c Z^{bc}_{OA}}{V^{c*}_{A}} \\ 
                                
                               \frac{ \Delta S_{A}^a Z^{ca}_{OA}}{V^{a*}_{A}}+\frac{ \Delta S_{A}^b Z^{cb}_{OA}}{V^{b*}_{A}} + \frac{ \Delta S_{A}^c Z^{cc}_{OA}}{V^{c*}_{A}} \\ 
                                
                               \end{bmatrix}
\end{equation}
where $\vect{\Delta V_O}$ is a vector consisting of the voltage change in phases a, b, and c at an observation node $O$ given by $\Delta V^a_O$, $\Delta V^b_O$ and $\Delta V^c_O$.  $V^*_{A}$ and $\Delta S_{A}$ represent the complex conjugate of voltage and complex power change at actor node $A$, respectively. The superscripts a, b, and c represent different phases and $Z$ corresponds to the self and mutual impedance of the shared line between the actor and observation node. The voltage change due to multiple actor nodes $A \in \mathcal{A}$ can be formulated as the cumulative effect of all actor nodes on a particular observation node as given in corollary 1 \cite{Munikoti2020probabilistic}.\\

\textbf{Corollary 1.} For a given three phase distribution network, the cumulative change in complex voltage at an observation node $O$ due to the change in complex power at multiple actor nodes can be formulated as:
\begin{equation}  
  \vect{\Delta V_O} \approx   - \sum_{A \in \mathcal{A}}\begin{bmatrix} \frac{ \Delta S_{A}^a Z^{aa}_{OA}}{V^{a*}_{A}}+\frac{ \Delta S_{A}^b Z^{ab}_{OA}}{V^{b*}_{A}} + \frac{ \Delta S_{A}^c Z^{ac}_{OA}}{V^{c*}_{A}} \\
 
                               \frac{ \Delta S_{A}^b Z^{ba}_{OA}}{V^{a*}_{A}}+\frac{ \Delta S_{A}^b Z^{bb}_{OA}}{V^{b*}_{A}} + \frac{ \Delta S_{A}^c Z^{bc}_{OA}}{V^{c*}_{A}} \\ 
                                
                               \frac{ \Delta S_{A}^a Z^{ca}_{OA}}{V^{a*}_{A}}+\frac{ \Delta S_{A}^b Z^{cb}_{OA}}{V^{b*}_{A}} + \frac{ \Delta S_{A}^c Z^{cc}_{OA}}{V^{c*}_{A}} \\ 
                                
                               \end{bmatrix}
\end{equation}
where, $\mathcal{A}$ represents the set of all actor nodes resulting in the complex voltage change at node $O$. The analytical method presented in corollary 1 gives us a computationally efficient method for computing the probability of voltage change at any given observation node $O$ due to change in complex power at multiple actor nodes $A \in \mathcal{A}$. Further, the execution time of the method to calculate the voltage sensitivity for a single observation node is an order faster (e.g., with an intel i7 processor based PC, it is 0.00871s, compared to 0.0537s in classical load flow method for the modified IEEE 37 bus system). This clearly shows that the proposed approach has an edge over traditional methods in terms of computational efficiency and the difference further increases with the size of the network. The analysis in this paper is based on the probabilistic extension of corollary 1. 
\section{Preemptive Voltage Violation Prediction}
\label{section:preemptive}
The voltage sensitivity analysis derived in section II is extended to predict the probability distribution of voltage at an observation node due to complex power change at multiple actor nodes. The analytical approach in this work assumes that based on measurements of complex power and voltages at a subset of locations, it is possible to estimate voltage states across the entire network, similar to the approaches presented in \cite{jhala2017probabilistic} \cite{jhala2019data}. The variability in complex power injection or consumption at actor nodes results in random voltage fluctuations at observation nodes. In this case, actor nodes represent active consumers integrated with distributed PV generation. Subsequently, if $\vect{V^p_O}$ is the present three phase voltage at an observation node $O$ that is obtained from system measurements, then $\vect{V^f_O}$ represents the future predicted complex voltage vector at that particular observation node. $\vect{V^f_O}$ is expected to be random due to the uncertainty introduced by the distributed PV generation and corresponds to,
  \begin{equation}  
 \vect{V^f_O} = \vect{V^p_O} + \vect{\Delta V_O}.
\end{equation}
Here, $\vect{\Delta V_O}$ represents the change in voltage at an observation node due to random complex power changes at actor nodes. Considering a single phase for simplicity, the voltage change at an observation node $O$ due to single actor node $A$ can be expressed in terms of real and imaginary part of voltage change as follows:
  \begin{equation}  
 \Delta V^a_{OA} = \Delta V^{a,r}_{OA} +j \Delta V^{a,i}_{OA}
 \label{eq:change}
\end{equation}
where,
\begin{multline}
 \Delta V^{a,r}_{OA} = -\frac{1}{|V^a_A|}(\Delta P^a_A(R^u_{OA} \cos \theta_A - X^u_{OA} \sin \theta_A)\\
 - \Delta Q^a_A(R^u_{OA} \sin \theta_A + X^u_{OA} \cos \theta_A))
\end{multline}
and, 
\begin{multline}
  \Delta V^{a,i}_{OA} = -\frac{1}{|V^a_A|}(\Delta Q^a_A(R^u_{OA} \cos \theta_A - X^u_{OA} \sin \theta_A)\\
 + \Delta P^a_A(R^u_{OA} \sin \theta_A + X^u_{OA} \cos \theta_A))
\end{multline}
where, $u$ represents different phase sequences, i.e., aa, ab, ac in phase a. $\Delta P^a_A$ and $\Delta Q^a_A$ represent the active and reactive power changes at phase a of actor node $A$, $R^u_{OA}$ and $X^u_{OA}$ are the real and imaginary parts of the impedance of the shared line between the observation $O$ and actor node $A$, and $\theta_A$ is the phase angle of the voltage at the actor node $A$.\\
Similar to corollary 1, (\ref{eq:change}) can be extended to accommodate the impact of multiple actor nodes. Therefore, the cumulative voltage change at a single phase in an observation node $O$ due to multiple actor nodes $A \in \mathcal{A}$ can be written as:
  \begin{equation}  
 \Delta V_O = \sum_{A \in \mathcal{A}}\Delta V^a_{OA} = \sum_{A \in \mathcal{A}}\Delta V^{a,r}_{OA} +j \sum_{A \in \mathcal{A}}\Delta V^{a,i}_{OA}
\end{equation}
At each time instant, the active and reactive power injections in the system can be modeled as random variables based on the variability of distributed PV generation at active consumer sites. Therefore, it is natural to model $\Delta V_O$ as a random variable as well. The derivation of the distribution of $|\Delta V_O|$ is the focus of the next subsection.
\subsection{Probability Distribution of predicted voltage}
Theorem 2 provides the probability distribution of the magnitude of predicted voltage at an observation node $O$ due to complex power change at multiple actor nodes $A \in \mathcal{A}$ for a single phase.\\
\textbf{Theorem 2.} For a given unbalanced distribution network, the predicted voltage magnitude ($|V^f_O|$) at an observation node $O$  due to complex power changes at multiple actor nodes $A \in \mathcal{A}$ follows a $Rician$ distribution, i.e.,
  \begin{equation}  
  |V^f_O|  \sim  Rician( \kappa, \sigma)
  \label{eq:ricianbeforeproof}
\end{equation}
where, $\kappa = \sqrt{w}$ and $\sigma = \sqrt{\lambda}$ with, 
  \begin{equation}  
  \lambda  =  \frac{\sigma^4_r(1+2\mu^2_r)+\sigma^4_i(1+2\mu_i^2)}{\sigma^2_r(1+2\mu_r^2)+\sigma^2_i(1+2\mu_i^2)}
\end{equation}
and, 
  \begin{equation}  
  w =  \frac{(\sigma^2_r \mu_r^2+\sigma^2_i \mu_i^2)(\sigma^2_r + \sigma^2_i +2 \sigma^2_r \mu_r^2 +2 \sigma^2_i \mu_i^2)}{\sigma_r^4 + \sigma_i^4 + 2 \sigma_r^4 \mu_r^2 + 2 \sigma_r^4 \mu_i^2}
\end{equation}
here,  $\sigma^2_r = \vect{c_r}^T \Sigma_{\Delta S} \vect{c_r}$, $\sigma^2_i = \vect{c_i^T} \Sigma_{\Delta S} \vect{c_i}$, $\mu_r=V_O^{r,p}+\vect{c_r^{T}}\vect{\mu_{\Delta S}}$, and $\mu_i=V_O^{i,p}+\vect{c_i^{T}}\vect{\mu_{\Delta S}}$. In this context, $V^{r,p}_O$ and $V^{i,p}_O$ are the present estimated values of real and imaginary parts of voltage. $\vect{c_r}$ and $\vect{c_i}$ are based on system topology and $\vect{\mu_{\Delta S}}$ and $\Sigma_{\Delta S}$ are related to variability in power change as will be discussed in the proof. \\
\textbf{Proof.} The variability of PV generation randomizes the associated power output. In this case, the forecasted power change is modeled as a non-zero mean random vector with mean $\vect{\mu_{\Delta S}}$ and covariance $\Sigma_{\Delta S}$. This model captures a nominal forecast ($\vect{\mu_{\Delta S}}$) and the associated error in forecast characterized by $\Sigma_{\Delta S}$.  The real and reactive power represent the net nodal load changes given the presence of distributed PV generation at active consumer sites. Accordingly,  $\vect{\Delta S}$ can be represented as shown in (\ref{eq:deltas}) with $n$ representing the number of nodes in the system.
  \begin{equation}  
\vect{\Delta S} = [\Delta P_1^a, . . ., \Delta P_n^a, \Delta Q_1^a, . . ., \Delta Q_n^a]
\label{eq:deltas}
\end{equation}
The following steps detail the steps involved in the derivation of the distribution of $|V^f_O|$.\\
\subsubsection{Computation of covariance matrix $\Sigma_{\Delta S}$}
The covariance matrix $\Sigma_{\Delta S}$ captures the relationship between complex power changes at multiple actor nodes and can be determined based on historical measurements. For a given system, the diagonal elements of the covariance matrix (i.e., variance) depend on the size of distributed PV generation and the uncertainty in the forecast. The off diagonal elements of the covariance matrix are based on the future net-load forecasts given a particular spatial PV generation and load profile. If a particular node in the network is not integrated with distributed PV generation, then the mean and variance term of the respective node is equivalent to their typical load variability. Accordingly, the covariance matrix $\Sigma_{\Delta S}$ can be formulated as,
\begin{equation}
\resizebox{1.008\hsize}{!}{$\Sigma_{\Delta S} = \begin{bmatrix}
   \sigma^2_{p_1} & \dots & cov(p_n,p_1) & cov(q_1,p_1) & \dots & cov(q_n,p_1) \\
    \vdots & \ddots & \vdots & \vdots & \ddots & \vdots \\
    cov(p_1,p_n) & \dots & \sigma^2_{p_n} & cov(q_1,p_n) & \dots & cov(q_n,q_n) \\
    cov(p_1,q_1) & \dots & cov(p_n,q_1) & \sigma^2_{q_1} & \dots & cov(q_n,p_1) \\
    \vdots & \ddots & \vdots & \vdots & \ddots & \vdots \\
    cov(p_1,q_n) & \dots & cov(p_n,q_n) & cov(q_1,q_n) & \dots & \sigma^2_{q_n} \\
    \end{bmatrix}$}
\end{equation}
Here, $n$ represents the number of nodes in the desired network and $p_i$ and $q_i$ are the active and reactive power injection or consumption at the $i^{th}$ active consumer site, respectively. $\sigma^2_{p_i}$ and $\sigma^2_{q_i}$ capture the variance of active and reactive power generation across different actor nodes, respectively, and the off diagonal elements capture the correlation between various generators due to geographical proximity. 
\subsubsection{Computation of \boldmath{$c_r$} and \boldmath{$c_i$} vectors}
The present work assumes prior knowledge of the system parameters. To begin with, define $\vect{c_r}$ and $\vect{c_i}$ as follows:
\begin{equation}
\vect{c_r}=[\vect{c^{aa}_r}, \vect{c^{ab}_r}, \vect{c^{ac}_r}]^T,
\vect{c_i}=[\vect{c^{aa}_i}, \vect{c^{ab}_i}, \vect{c^{ac}_i}]^T
\label{eq:CRCIvectors}
\end{equation}
For simplicity, the vectors are shown for single phase, i.e., phase a, where each vector is composed of three sub-vectors corresponding to self and mutual phases. $\vect{c_r}$and $\vect{c_i}$  for a single phase can be computed as,
\begin{equation}
\vect{c^{aa}_r} = \begin{bmatrix}
   \frac{-(R^{aa}_{O1}\cos(\theta_1)-X^{aa}_{O1}\sin (\theta_1) )}{|V^a_1|} \\
    \vdots  \\
   \frac{-(R^{aa}_{On}\cos(\theta_n)-X^{aa}_{On}\sin (\theta_n) )}{|V^a_n|}  \\
   \frac{(R^{aa}_{O1}\sin(\theta_1)+X^{aa}_{O1}\cos (\theta_1) )}{|V^a_1|}  \\
    \vdots  \\
   \frac{(R^{aa}_{On}\sin(\theta_n)+X^{aa}_{On}\cos (\theta_n) )}{|V^a_n|}   \\
    \end{bmatrix}
\end{equation}
\begin{equation}
\vect{c^{aa}_i} = \begin{bmatrix}
   \frac{-(R^{aa}_{O1}\sin(\theta_1)+X^{aa}_{O1}\cos (\theta_1) )}{|V^a_1|} \\
    \vdots  \\
   \frac{-(R^{aa}_{On}\sin(\theta_n)+X^{aa}_{On}\cos (\theta_n) )}{|V^a_n|}  \\
   \frac{-(R^{aa}_{O1}\cos(\theta_1)-X^{aa}_{O1}\sin (\theta_1) )}{|V^a_1|}  \\
    \vdots  \\
   \frac{-(R^{aa}_{On}\cos(\theta_n)-X^{aa}_{On}\sin (\theta_n) )}{|V^a_n|}   \\
    \end{bmatrix}
\end{equation}
The aforementioned vectors are constant for a given system with a particular set of active consumer (actor) nodes integrated with distributed PV generation. The elements of $\vect{c_r}$ and $\vect{c_i}$ vectors consist of the ratio of the impedance of shared path (between the observation and actor node) to the rated voltage of the associated phase, (e.g., in this case, it would be phase a). When the system topology changes, the $\vect{c_r}$ and $\vect{c_i}$ vectors are expected to change as well. 
\subsubsection{Probability distribution of $\Delta V_o^r$ and $\Delta V_o^i$}
This subsection provides an expression for the real and imaginary parts of voltage change at an observation node due to complex power change at multiple actor nodes. The change in voltage at an observation node is expressed as the sum of voltage changes induced by each actor node as shown by corollary 1 in section II. Thus, the probability distribution of real and imaginary part of voltage change are formulated as follows:
\begin{equation}
\Delta V_O^{a,r} = \sum_{A \in \mathcal{A}} \Delta V^r_{OA} = \vect{c_r^T} \vect{\Delta S} \overset{D}{\rightarrow} \mathcal{N}(\vect{c^T_{r}}\vect{\mu_{\Delta S}}, \vect{c_r^T}\Sigma_{\Delta S} \vect{c_r})
\label{eq:clt1}
\end{equation}
\begin{equation}
\Delta V_O^{a,i} = \sum_{A \in \mathcal{A}} \Delta V^i_{OA }= \vect{c_i^T} \vect{\Delta S} \overset{D}{\rightarrow} \mathcal{N}(\vect{c^T_{i}}\vect{\mu_{\Delta S}}, \vect{c_i^T}\Sigma_{\Delta S} \vect{c_i})
\label{eq:clt2}
\end{equation}
where $\mathcal{A}$ represents the set of actor nodes resulting in voltage change at the observation node $O$. Using Lindeberg-Feller CLT, (\ref{eq:clt1}) and (\ref{eq:clt2}) indicate that $\Delta V_O^{a,r}$ and $\Delta V_O^{a,i}$ converge in distribution to a Gaussian random variable.\\
The covariance between real $\Delta V_O^r$ and imaginary $\Delta V_O^i$ parts of voltage change corresponds to $cov(\Delta V_O^r,\Delta V_O^i)=\vect{c_r^T}\Sigma_{\Delta S} \vect{c_i}$. Thus, the real and imaginary parts of voltage change at an observation node can be rewritten as a multi-variate normal vector corresponding to,
\begin{equation}
\vect{\Delta V_O} \triangleq 
 \begin{bmatrix}
    \Delta V_O^r \\
    \Delta V_O^i \\
    \end{bmatrix}
\sim \mathcal{N}(\vect{\mu_1}, \Sigma_1)
\end{equation}
where,
\begin{equation}
\vect{\mu_1} = 
 \begin{bmatrix}
    \vect{c_r^T}\vect{\mu_{\Delta S}} \\
    \vect{c_i^T}\vect{\mu_{\Delta S}} \\
    \end{bmatrix}
\end{equation}
\begin{equation}
\Sigma_1 = 
 \begin{bmatrix}
    \vect{c_r^T}\Sigma_{\Delta S} \vect{c_r} &\vect{c_r^T}\Sigma_{\Delta S} \vect{c_i} \\
    \vect{c_r^T}\Sigma_{\Delta S} \vect{c_i} &\vect{c_i^T}\Sigma_{\Delta S} \vect{c_i}\\
    \end{bmatrix}
\end{equation}
Recall the expression of $\vect{V_O^f}$ and  $\vect{V_O^p}$ from section II. The real and imaginary parts of predicted voltage can be written as:
\begin{equation}
\vect{V^f_O} \triangleq 
 \begin{bmatrix}
   V_O^{r,f} \\
   V_O^{i,f} \\
    \end{bmatrix}
=  \begin{bmatrix}
    V_O^{r,p} \\
    V_O^{i,p} \\
    \end{bmatrix}
    + \begin{bmatrix}
    \Delta V_O^{r} \\
    \Delta V_O^{i} \\
    \end{bmatrix}
\end{equation}
\begin{equation}
\vect{V_O^f} \sim \mathcal{N}( \begin{bmatrix}                             V_O^{r,p}+\vect{c_r^{T}}\vect{\mu_{\Delta S}}\\
 
                       V_O^{i,p}+\vect{c_i^{T}}\vect{\mu_{\Delta S}}  \\ 
                                
                               \end{bmatrix}, \Sigma_1 )
\end{equation}

\begin{equation}
|V_O^f|^2 = (V_O^{r,f})^2+(V_O^{i,f})^2
\end{equation}

\begin{figure}[t!]
\includegraphics[width=8.5cm]{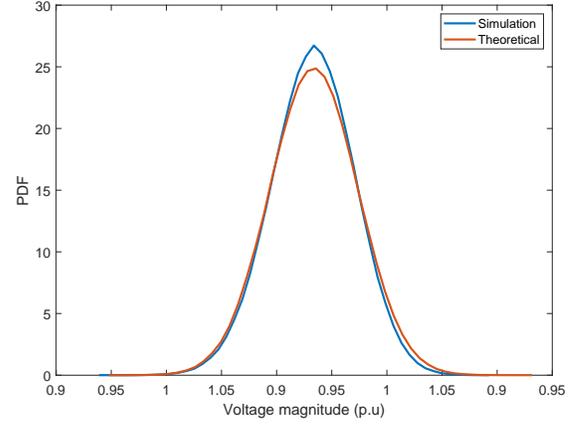}
\centering
\caption{Predicted voltage using (\ref{eq:rice}) vs. load flow.}
  \label{fig:validation}
\end{figure}

The distribution of the squared magnitude of $V^f_O$ is a sum of dependent non-central chi-square distributions. Each real and imaginary part of the predicted voltage follows non zero mean Gaussian distribution and thus their squares will have a non central chi square distribution \cite{mathai1992quadratic}, 
\begin{equation}
|V_O^f|^2 \sim \sigma_r^2 \chi_1^2(\mu_r^2)+\sigma_i^2 \chi_1^2(\mu_i^2)
\end{equation}
where, $\sigma$ and $\mu$ are the weight and non centrality parameter of non central chi square distribution with one degree of freedom, respectively. The sum of weighted non-central chi-square distributions can then be approximated with a scaled non-central chi-square with weight $\lambda$, non-centrality parameter $w$, and $v$ degrees of freedom as\cite{mathai1992quadratic}:
\begin{equation}
|V_O^f|^2 \sim \lambda \chi_v^2(w)
\end{equation}
where,
\begin{equation}  
 \lambda  =  \frac{\sigma^4_r(1+2\mu^2_r)+\sigma^4_i(1+2\mu_i^2)}  {\sigma^2_r(1+2\mu_r^2)+\sigma^2_i(1+2\mu_i^2)}
\end{equation}
\begin{equation}  
  w =  \frac{(\sigma^2_r \mu_r^2+\sigma^2_i \mu_i^2)(\sigma^2_r + \sigma^2_i +2 \sigma^2_r \mu_r^2 +2 \sigma^2_i \mu_i^2)}{\sigma_r^4 + \sigma_i^4 + 2 \sigma_r^4 \mu_r^2 + 2 \sigma_r^4 \mu_i^2}
\end{equation}
\begin{equation}  
  v =  \frac{(\sigma^2_r +\sigma^2_i)(\sigma^2_r +\sigma^2_i + 2 \sigma^2_r \mu_r^2 +2 \sigma^2_i \mu_i^2)}{\sigma^2_r +\sigma^2_i + 2 (\sigma^4_r \mu_r^2)+2 (\sigma^4_i \mu_i^2)}
\end{equation}
Since the square root of  a non-central chi-square random variables follows a Rician distribution \cite{mathai1992quadratic}, the magnitude of predicted voltage change will follow a Rician distribution:
  \begin{equation}  
  |V^f_O|  \sim  Rician(\kappa, \sigma)
  \label{eq:rice} 
\end{equation}
where, $\kappa = \sqrt{w}$ and $\sigma = \sqrt{\lambda}$, which is consistent with (\ref{eq:ricianbeforeproof}). This expression is first validated on the modified IEEE 37 bus system. Figure \ref{fig:validation} shows the predicted voltage at observation node 22 using the expression derived in (\ref{eq:rice}) vs. the values calculated using the load flow method. For the current setup, four arbitrary actor nodes are chosen for the validation test, namely, 2, 11, 20, and 29. The Jensen-Shannon distance between the theoretical and simulated distribution is in the order of $10^{-2}$. Jensen-Shannon distance ranges from 0 to 1 indicating exact distribution match and mismatch, respectively. Thus, the proposed method is highly accurate in predicting the distribution of voltage at a particular observation node. 
\subsection{Assessment of node vulnerability to voltage violation}
\label{section:assessment}
The aim of this work is to identify nodes with high probability of voltage violation. The expression derived in (\ref{eq:rice}) shows that the predicted voltage magnitude $ |V^f_O| $ follows a Rician distribution. $P_v(t)$, the probability of node voltage violation at a given time instant corresponds to
  \begin{equation}  
  P_v(t) = 1 - P(0.95 < |V^f_O| < 1.05).
  \label{eq:criteria}
\end{equation}
(\ref{eq:criteria}) can be used to identify vulnerable nodes by comparing $P_v(t)$ with a particular threshold. The threshold used in this paper is 0.5, i.e., nodes with voltage-violation probabilities higher than 0.5 are considered vulnerable. The method is generic and can be implemented on all observation nodes in the network. This assessment provides an insight into the voltage status of the network at a future time instant. The assessment criterion is computationally efficient and the outcome can be used as an input for voltage control. The control aspects, although not discussed in this paper, will be part of future research efforts. 
\section{Simulation and results}
\begin{figure}[t!]
\includegraphics[width=8.5cm]{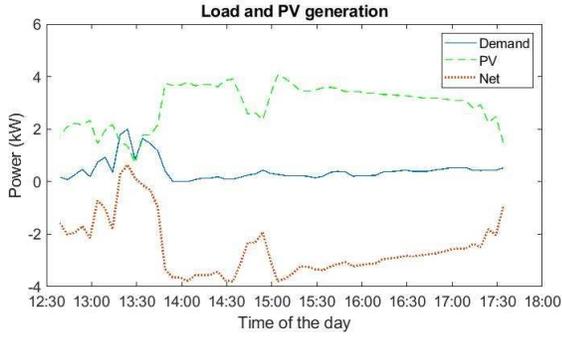}
\centering
\caption{Solar PV generation profile for each unit.}
\label{fig:pvprofile}
\end{figure}

\begin{figure}[t!]
\includegraphics[width=8.5cm]{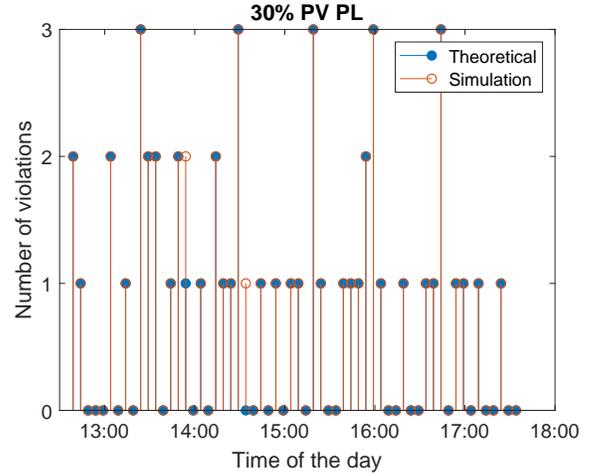}
\centering
\caption{Voltage violation prediction using (\ref{eq:rice}) vs. load flow.}
\label{fig:30pl}
\end{figure}

\begin{figure}[t!]
\includegraphics[width=8.5cm]{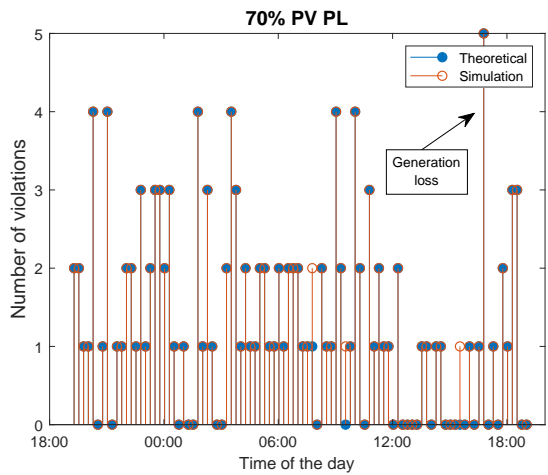}
\centering
\caption{Voltage violation prediction using (\ref{eq:rice}) vs. load flow.}
\label{fig:genloss}
\end{figure}

This section summarizes the simulation results and findings related to PVSA based preemptive voltage monitoring strategy. First, the violation prediction method is tested on the IEEE 37 node test system. Next, a catastrophic scenario is presented where the system experiences a complete loss of generation at a particular actor node and the efficacy of the proposed voltage violation prediction is evaluated. 
 Actual voltage violations in the system are extracted using power flow solutions for the purpose of validating the proposed approach. Among the 37 system nodes, a subset of nodes is considered to be active consumers with integrated distributed PV generation and voltage status is monitored on all system (observation) nodes. For the first case, A hypothetical solar PV generation scenario is considered from noon to 18:00 with power and voltage measurement availability every 15 minutes. The solar PV generation in this work is modeled as a random process with a component of uncertainty to illustrate a profile that follows real world scenarios as follows: 
  \begin{equation}  
  G_{PV}(t) = S(t)+R_s(t).
\end{equation}
Here, $S(t)$ is the mean forecast trend of the solar PV generation and $R_s(t)$ represents a zero mean uncorrelated Gaussian random process illustrating the uncertainty in PV generation. Figure \ref{fig:pvprofile} shows the solar PV generation model used ($S(t)$) as well as the net-power curve used for simulation in this paper. Time instances where the net-power is negative indicates reversed power flow in the grid due to surplus solar PV generation. Although this particular scenario is considered, the proposed method is generic and applicable to different scenarios. 
Initially, net-power injections, system data such as node locations and line impedances are used to compute the cumulative effect of actor nodes on all observation nodes in the network. The analytical expressions presented in section \ref{section:background} are the basis for estimating the mean and variance of voltage at all nodes as discussed in section \ref{section:preemptive}. The covariance matrix $\Sigma_{\Delta S}$ is computed relying on estimates of historical data. The network topology is used to compute vectors $\vect{c_r}$ and $\vect{c_i}$ as formulated in (\ref{eq:CRCIvectors}). Finally, node voltage state estimates as well as the analytical voltage change probability distribution are utilized to compute the probability of node voltage violation according to the threshold given in section \ref{section:assessment}. In the first case study, a hypothetical 30\% Penetration Level (PL) of distributed PV generation is randomly allocated among 14 actor nodes and voltage state is monitored across all observation nodes in the network. Figure \ref{fig:30pl} shows the number of violations in the system using the proposed method in (\ref{eq:rice}) vs. load flow method. From figure \ref{fig:30pl}, it can be inferred that the proposed method accurately predicts voltage violations in the system compared to actual violations calculated using load flow method.\\
In the  next case study, a scenario with complete loss of PV generation at a certain time instant is investigated. In this case, actor nodes are assigned to three different 24 hour PV generation profiles contributing to a 70 \% PL for demonstrating the generality of the proposed method. The system in this case consists of 20 arbitrary active consumer nodes integrated with distributed PV generation. Similar to the first case study, voltage state is monitored across all nodes in the network. Figure \ref{fig:genloss} shows the number of voltage violations in the system using the proposed analytical method in (\ref{eq:rice}) vs. load flow method with a PV generation loss scenario occurring at time 16:32 of the day. It can be inferred that the proposed method effectively predicts voltage violations not only under normal operation conditions but also under generation loss scenarios. 
Finally, the accuracy of the proposed method is quantified via multiple Monte-Carlo simulations. Two cases are considered for investigating the accuracy of the proposed method, namely, 30\% and 70\% PLs. For both cases, 20 arbitrary actor nodes are integrated with distributed PV generation and voltage state is monitored across all nodes in the network. Both scenarios are simulated for 100 Monte-Carlo simulations and the mean prediction error  is obtained. Table \ref{table:error} shows the prediction error for both cases, which demonstrates that the effectiveness of the proposed method in predicting voltage violations in the system is higher than 95\%. Therefore, the proposed method can provide effective foresight on voltage violations to system operators, which can then be utilized to implement an appropriate optimal voltage control strategy.
\begin{table}[th!]
\caption{Theoretical vs. actual voltage violations.}
\centering
\begin{tabular}{{|P{2.6cm}|P{2.6cm}|P{2.6cm}|}}
\hline
PL    & Prediction error ($\%$) \\ \hline
30\%                    &  4.31              \\ \hline
70\%                    & 4.43              \\ \hline
\end{tabular}
\label{table:error}
\end{table}
\section{Conclusion}
Power systems across the globe are witnessing rapid integration of smart grid technologies including  renewable energy based distributed generation. This increased integration increases system vulnerability to voltage violations which greatly decreases system reliability. Conventional voltage control methods rely mainly on reactionary methods which makes it difficult to completely mitigate voltage violations in the system. This paper proposes a new preemptive voltage monitoring method that provides useful foresight on violations in the system. The proposed approach is based on probabilistic voltage sensitivity analysis where the probability of voltage violation is computed for all system nodes given changes in power injections at different system nodes. Results in this paper demonstrate that the proposed voltage violation prediction method is extremely accurate with a low prediction error of approximately 4 \%. \\
Future research directions of this work include identifying the most dominant and influencial nodes that result in voltage violations at an observation node which in turn can be used to develop quick and effective control solutions.
\section{ACKNOWLEDGMENT}
This material is based upon work partly supported by the Department of Energy, Office of Energy Efficiency and Renewable Energy (EERE), Solar Energy Technologies Office, under Award \# DE-EE0008767 and National science foundation under award \# 1855216.
\bibliographystyle{IEEEtran}
\bibliography{references}
\end{document}